\documentclass[12pt,preprint]{aastex}

\usepackage{amsmath}

\slugcomment{To appear in PASP}

\shorttitle{$uvby$ Passbands and Zeropoints}
\shortauthors{Bessell}

\begin{document}


\title{Photonic Passbands and Zeropoints for the Str\"{o}mgren $uvby$ System. }


\author{Michael S. Bessell }
\affil{RSAA, Mount Stromlo Observatory, The Australian National University, ACT 2611, Australia}
\email{bessell@mso.anu.edu.au}

\begin{abstract}
Photonic passbands have been derived for the $uvby$ standard system by convolving the original filter passbands of Str\"{o}mgren and Perry with atmospheric extinction and the QE of a cooled 1P21 photomultiplier tube. Using these new passbands, synthetic photometry was calculated for all the stars in the extensive NGSL and MILES spectrophotometric libraries  and compared with the homogenised $b-y$, $m_{1}$ and $c_{1}$ indices in the Hauck-Mermilliod 1998 catalog and the derived $u-v$ and $v-b$ colors.  Excellent agreement between observed and synthetic photometry was achieved with regression slopes near unity. Slightly better fits were obtained by considering stars with $b-y < $ 0.5 and $b-y > $ 0.5, separately. It is recommended that these new passbands be used together with the provided transformation equations to generate synthetic photometry from model atmosphere fluxes and observed spectrophotometry. 
Synthetic photometry was also carried out using the natural system of the 4-channel spectrograph-photometers  and those of Cousins and Eggen in order to explore the systematic differences that could be expected between their instrumental systems and the standard system. 
\end{abstract}


\section{Introduction}
The Str\"{o}mgren $uvby$ system \citep{Strom66}, is a widely used  intermediate-band photometric system comprising data for more than 63,300 stars in the homogenised \citet{Hauc98} catalog. It has long been considered a precise and accurate standard system that is very useful for determining temperatures and effective gravities of B, A, F and early- G stars and metallicities of F and G stars. However, the use of different filters at many observatories and the extension of the photometry to late-G and K stars and to supergiants and reddened stars has resulted in significant systematic differences in the instrumental systems between observers. The recommended method of photometric data reduction to derive the indices $c_{1} = (u-v)-(v-b)$ and $m_{1}= (v-b)-(b-y)$, rather than the colors $u-v$, $v-b$ directly, has also resulted in larger uncertainties in the transformed indices, mainly for the cooler stars. These problems have been vividly illustrated and discussed by \citet{Manf84,Manf87}. 

 A large database of excellent $uvby$ photometry has been obtained with the four 4-channel spectrograph-photometers \citep{Helt87} on telescopes at Kitt Peak, La Silla and San Pedro Martir \citep[eg.][]{Olse83, Olse93, Schu88}. Fig 1 in \citet{Olse83} shows the complicated decision processes carried out to transform this well defined and stable natural system photometry onto the standard system. \citet{Cous87} also discussed the non-linearities involved in successively transforming his E-region photometry onto the standard system and \citet{Egge76} outlined issues associated with too-narrow  a $v$ filter. Given these different and complex transformations one may wonder whether passbands can be defined that accurately represent the homogenised $uvby$ system. 
 
In the last few years, two libraries of accurate higher resolution (R$\sim$1000-2000) spectrophotometric data have become available -  NGSL \citep[][\url{http://archive.stsci.edu/prepds/stisngsl/index.html}]{Heap07} and MILES \citep[][\url{http://www.iac.es/proyecto/miles/}]{Sanc06} and most of the stars in these spectral libraries also have Hipparcos $H_{p}$ magnitudes - enabling \citet{BesM11} to renormalize the spectrophotometry on to a precise absolute scale by synthesizing the $H_{p}$ magnitudes. Many of these stars are also in the \citet{Hauc98} $uvby$ catalog thus providing the opportunity to critically compare the synthetic $uvby$ photometry with the standard values.
  
 \section{The $uvby$ passbands}
\citet{Mats69} and  \citet{Craw79} published transmissions of the KPNO $uvby$ filter set No.1 that was used when setting up the original $uvby$ system by \citet{StromP65}. These filter functions are considered to be the best starting point for synthetic $uvby$ photometry calculations. Synthetic photometry based on model atmosphere fluxes have been computed by many authors \citep{Mats69, Rely78, Kuru79, Clem04, Oneh09} and while most have convolved the filter functions with a 1P21 sensitivity function, Al mirror reflectivity and 1 airmass of extinction,  some \citep[eg][]{Mats69, Maiz06} have used the filter transmissions directly. 

The product of the filter passband, the atmospheric transmission, the mirror reflectivity and the detector sensitivity function is called the system response function or system passband. In the past, the detector sensitivity function that was used related to the energy measured. Nowadays with photon counting detectors being the norm, the detector sensitivity function used relates to the number of photons measured. The resultant system response functions or system passbands therefore differ depending whether they reflect the energy of the photons or the number of photons. As modern synthetic photometry packages, such as $\it{synphot}$ assume that system response functions are in the photon form, we will use that form in this paper and also use the terms photonic response function or photonic passband to distinguish them from the alternative energy forms.

The 1P21 sensitivity function is from \citet[][table 7]{Kuru79}; this has the same UV-cuton as measured by \citet[][table1.1]{Bess79} and appears to have a similar red cutoff to that measured by \citet[][fig 6]{Youn63}. It is presumed to be in terms of the photocathode radiant response (in units of mA W$^{-1}$) and we have converted it into QE to derive the photonic responses \citep[see ][]{BesM11}. The mirror reflectivity \citep[see values in][]{Helt87} was neglected, but we applied 1.2 air masses of typical Siding Spring Observatory (1200m) extinction.

In Table 1 we list our adopted normalized system photonic passbands $S_{x} (\lambda)$ for $u$, $v$, $b$ and $y$, and in Fig 1 we  show the passbands compared to those of  \citet{Maiz06} and the 4-channel spectrograph-photometers \citep{Helt87} (see below). 

\newcommand{\bigint}{\displaystyle \int } 
\section{Comparison between synthetic photometry and standard photometric values}
The synthetic photometry was computed by evaluating, for each of the $uvby$ bands, the expression \citep[see][]{BesM11} 
\begin{equation} 
$$ mag$_{x}$ = AB$_{\nu}$ $-$ ZP$_{x}$   $$
\end{equation}
where 
\begin{equation}
$$ AB =   $-2.5 \log\frac{\bigint{f_{\nu}(\nu) S_{x}(\nu)  d\nu/\nu}} {\bigint{S_{x}(\nu) d\nu/\nu}} - 48.60$ =  $-2.5 \log\frac{\bigint{f_{\lambda}(\lambda) S_{x}(\lambda) \lambda d\lambda}} { \bigint{S_{x}(\lambda)c\,d\lambda /{\lambda}}} - 48.60$     $$
\end{equation}
and $f_{\nu}(\nu)$ is the observed absolute flux in erg cm$^{-2}$ sec$^{-1}$ Hz$^{-1}$, $f_{\lambda}(\lambda)$ is the observed absolute flux in erg cm$^{-2}$ sec$^{-1}$ \AA$^{-1}$,  $S_{x}(\lambda)$, is the photonic passband (Table 1), $\lambda$ is the wavelength in \AA, and ZP$_{x}$ is the zeropoint magnitude for each band.

For accurate synthetic photometry it is important that the passbands provided to the integration routines are well sampled and smooth.  It is necessary therefore, to interpolate the coarsely sampled values of the passbands in Table 1 to a finer spacing of a few \AA\ using a univariate spline or a parabolic interpolation routine. 

Firstly, the CALSPEC stis005 spectrum for Vega was used to derive the AB$_{\nu}$ mag zeropoints of $-$0.308, $-$0.327, $-$0.187 and $-$0.027 for $u$, $v$, $b$ and $y$, respectively, adopting for Vega, $y$=0.03, \citep{Bess83}; $(b-y)$=0.004, $m_{1}$=0.157 and $c_{1}$=1.088, \citep{Hauc98}. 

We then computed synthetic colors for all the renormalized NGSL (376) and MILES (830) spectra \citep{BesM11} and matched the stars against the homogenised \citet{Hauc98} catalog, yielding 259 and 535 objects, respectively, with both spectrophotometry and $uvby$ photometry. 

The color differences and index differences were regressed against $b-y$, $m_{1}$ and $c_{1}$, and little evidence of residuals that were a continuous function of color were found. Any variation with color that was evident was better dealt with by splitting the sample into blue and red stars, rather than fitting an overall color term. Those stars with $b-y >$ 0.5  also showed slightly more scatter than did the bluer stars which could have been anticipated since the $uvby$ system was initially standardized for the bluer stars and extended later to the redder stars using a greater range of instrumental systems and only a few red standards.  (The higher scatter in the MILES comparison for those measures involving the $u$ band in this paper should be discounted because we extrapolated their UV fluxes somewhat crudely using model atmosphere fits to the rest of the spectrum.) Overall, it was extremely gratifying to see the near unity slopes and small scatter in the comparisons, indicating that the realised passbands are an excellent representation of the standard system. 

Figures 2 - 6 show the regressions for $b-y$, $m_{1}$, $c_{1}$, $u-v$ and $v-b$ overlayed with the lines fitted to the blue and red spectra separately. There were about 185 blue and 70 red spectra in the NGSL sample, and 360 blue and 150 red spectra in the MILES sample.
In many cases the same fitted line for blue and red stars could suffice within the uncertainties, but for some colors and indices, small blue/red star differences were evident. 

Figure 7 shows the synthetic $V-y$ regressions against $b-y$; the $V$ passband is from \citet{BesM11}. The mean wavelength  of the $y$ band is clearly very close to that of the $V$ band. 

In Table 2 are listed the results of the least-square linear fits to the regressions, including the uncertainties in the coefficients and the residual rms.   It is recommended that the transformation equations determined from the space-based NGSL spectra be used to convert synthetic photometry onto the same system as the \citet{Hauc98} catalog; however, the equations from the MILES fits are in excellent agreement with those from the NGSL spectra, although the scatter is slightly higher. It should also be noted that the slopes of the transformations are of the same order as the observational values reported by \citet{Craw79} using KPNO filter set No 1. 

\section{Review of some other natural $uvby$ systems}
\citet{Manf84} was the first to use synthetic photometry to explore the systematic effects that different instrumental system passbands have on $uvby$ photometry. This was further discussed by \citet{Manf87}. We decided to use the NGSL spectrophotometric data to examine two widely used natural $uvby$ systems, the 4-channel spectrograph-photometers \citep{Helt87} and that of \citet{Cous87}, in order to explore the systematic differences that could be expected between their instrumental systems and that presented in this paper as representing the standard system. The photonic passband of \citet{Helt87} are plotted in Fig 1 and listed in Table 3 (the narrow spike in the published Helt et al. $u$ passband has been smoothed over). 
 In Fig 8 and 9 we show the computed differences in the $u-v$ and $v-b$ colors plotted against $b-y$ for the NGSL spectra. The division between dwarfs and giants in these plots was made at log g $= 3.5$;  higher gravity stars are plotted as dwarfs and lower gravity as giants. The total range of the vertical scales of the two plots in each figure is the same.

The differences shown in the plots are very similar to those indicated by \citet{Olse83}, \citet{Cous87}, \citet{Manf84} and \citet{Manf87}. It is interesting to see the dwarf-giant separation for the cooler stars and the increase in scatter for the cooler giants. 

The effect of the narrow $v$ band ($\sim$110\AA\ fwhm) reported by \citet{Egge76} was also investigated. An Eggen-type $v$ band was constructed by simply scaling the half-width of the standard band and keeping the same central wavelength. In Fig 10 are shown the computed difference in the measured $v$ magnitude plotted against $b-y$ and $\beta$. Large difference are seen for the hotter stars due to the strength of the H$\delta$ line and in the cool giants and dwarfs due mainly to the atomic and molecular features. 

As pointed out by \citet{Manf87}, these comparisons underline the difficulties that observers with non-standard passbands experience in trying to standardize their internally precise $uvby$ photometry. 

\section{Summary} 
We have derived passbands for the $uvby$ system using the KPNO No.1 filter transmissions of \citet{Mats69} and \citet{Craw79} convolved  with the cathode sensitivity function of a 1P21 photomultiplier tube, and an extinction of 1.2 airmasses. These were converted to photonic passbands by dividing by the wavelength and then renormalized. They are listed in Table 1 and shown in Figure 1.

Initial AB mag $uvby$ zeropoints were derived from the stis005 spectrum of Vega.  Synthetic photometry was then carried out on the extensive NGSL and MILES spectrophotometric catalogs and compared with the observational data in the homogenised \citet{Hauc98} catalog. Excellent linear fits with near unity slopes were made to the synthetic-observational regressions showing how well the passbands represent the standard system. The coefficents of the fitted lines are given in Table 2 and the equations from the NGSL stars should be used to produce standard photometric values from synthetic photometry of observed spectrophotometric fluxes or model atmosphere fluxes. 

Some non-standard passbands of two well defined natural $uvby$ systems were also synthesized and shown to produce similar systematic differences to those reported by the users. These effects limit the accuracy with which transformations to the standard system can be made, mainly for the cooler stars and the hotter reddened stars.  The narrow $v$-band used by \citet{Egge76} and others  was shown to produce the largest systematic differences and supports Eggen's decision to not standardize his M1 and C1 photometry.  

However, in spite of these limitations, observers have generally been successful in transforming their photometry onto the standard system and \citet{Hauc98} have been successful in producing a very useful homogenised $uvby$ catalog. Hopefully, the revised passbands and transformation equations presented here will enable more reliable theoretical calibrations of Str\"{o}mgren indices to be made using model atmosphere fluxes. 

\acknowledgments
I wish to thank William Schuster for very helpful correspondence concerning the history and operation of the 4-channel spectrograph-photometers and the referee Chris Sterken for many suggestions that improved the paper. Vizier-R, Simbad, TopCat and Kaleidagraph were used in preparing this paper.

\begin{deluxetable}{cccccccc}
\tabletypesize{\scriptsize}
\tablenum{1}
\tablecaption{Normalized $uvby$ photonic passbands}
\tablewidth{0pt}
\tablehead{\colhead{Wave} & \colhead {$u$}&\colhead{Wave} & \colhead {$v$}& \colhead{Wave} & \colhead {$b$}& \colhead{Wave} & \colhead {$y$}}
\startdata
3150 & 0.000 & 3750 & 0.000 & 4350 & 0.000 & 5150 & 0.000 \\
3175 & 0.004 & 3775 & 0.003 & 4375 & 0.010 & 5175 & 0.022 \\
3200 & 0.050 & 3800 & 0.006 & 4400 & 0.023 & 5200 & 0.053 \\
3225 & 0.122 & 3825 & 0.016 & 4425 & 0.039 & 5225 & 0.082 \\
3250 & 0.219 & 3850 & 0.029 & 4450 & 0.056 & 5250 & 0.116 \\
3275 & 0.341 & 3875 & 0.044 & 4475 & 0.086 & 5275 & 0.194 \\
3300 & 0.479 & 3900 & 0.060 & 4500 & 0.118 & 5300 & 0.274 \\
3325 & 0.604 & 3925 & 0.096 & 4525 & 0.188 & 5325 & 0.393 \\
3350 & 0.710 & 3950 & 0.157 & 4550 & 0.287 & 5350 & 0.579 \\
3375 & 0.809 & 3975 & 0.262 & 4575 & 0.457 & 5375 & 0.782 \\
3400 & 0.886 & 4000 & 0.404 & 4600 & 0.681 & 5400 & 0.928 \\
3425 & 0.939 & 4025 & 0.605 & 4625 & 0.896 & 5425 & 0.985 \\
3450 & 0.976 & 4050 & 0.810 & 4650 & 0.998 & 5450 & 0.999 \\
3475 & 1.000 & 4075 & 0.958 & 4675 & 1.000 & 5475 & 1.000 \\
3500 & 0.995 & 4100 & 1.000 & 4700 & 0.942 & 5500 & 0.997 \\
3525 & 0.981 & 4125 & 0.973 & 4725 & 0.783 & 5525 & 0.938 \\
3550 & 0.943 & 4150 & 0.882 & 4750 & 0.558 & 5550 & 0.789 \\
3575 & 0.880 & 4175 & 0.755 & 4775 & 0.342 & 5575 & 0.574 \\
3600 & 0.782 & 4200 & 0.571 & 4800 & 0.211 & 5600 & 0.388 \\
3625 & 0.659 & 4225 & 0.366 & 4825 & 0.130 & 5625 & 0.232 \\
3650 & 0.525 & 4250 & 0.224 & 4850 & 0.072 & 5650 & 0.143 \\
3675 & 0.370 & 4275 & 0.134 & 4875 & 0.045 & 5675 & 0.090 \\
3700 & 0.246 & 4300 & 0.079 & 4900 & 0.027 & 5700 & 0.054 \\
3725 & 0.151 & 4325 & 0.053 & 4925 & 0.021 & 5725 & 0.031 \\
3750 & 0.071 & 4350 & 0.039 & 4950 & 0.015 & 5750 & 0.016 \\
3775 & 0.030 & 4375 & 0.027 & 4975 & 0.011 & 5775 & 0.010 \\
3800 & 0.014 & 4400 & 0.014 & 5000 & 0.007 & 5800 & 0.009 \\
3825 & 0.000 & 4425 & 0.006 & 5025 & 0.003 & 5825 & 0.004 \\
3850 & 0.000 & 4450 & 0.000 & 5050 & 0.000 & 5850 & 0.000 \\
\enddata
\end{deluxetable}

\begin{deluxetable}{rrrrrrrrrrr}
\tabletypesize{\scriptsize}
\tablenum{2}
\tablecaption{Zeropoints and slopes  in the form I$_{std}$ = a$_{0}$ + a$_{1}$I$_{syn}$ from synthetic photometry\tablenotemark{1}}
\tablewidth{0pt}
\tablehead{
\colhead{} & \colhead {}&\colhead{} & \colhead {NGSL\tablenotemark{2}}& \colhead{} & \colhead {}&  \colhead {}&\colhead{} & \colhead {MILES}& \colhead {}&\colhead{} \\
\colhead{index} & \colhead {a$_{0}$}&\colhead{$\pm$} & \colhead {a$_{1}$}& \colhead{$\pm$} & \colhead {rms}&  \colhead {a$_{0}$}&\colhead{$\pm$} & \colhead {a$_{1}$}& \colhead {$\pm$}&\colhead{rms}}
\startdata
$b-$y\tablenotemark{3} & $-$0.007 & 0.001 & 0.997 & 0.005 & 0.007 & 0.002 & 0.001 & 0.986 & 0.004 &  0.010  \\
       & 0.004 & 0.007 & 0.979 & 0.010 & 0.007 & 0.010 & 0.005 & 0.985 & 0.007  & 0.013   \\
\hline
$m_{1}$  & 0.005 & 0.002 & 0.963 & 0.010 & 0.022 & 0.021 & 0.002 & 0.988 & 0.012 & 0.019 \\
       & 0.011 & 0.005 & 0.951 & 0.012 & 0.022 & 0.015 & 0.004 & 0.979 & 0.010 & 0.026  \\
\hline
$c_{1}$  & $-$0.016 & 0.002 & 0.994 & 0.004 & 0.035 & $-$0.025: & 0.005 & 1.041: & 0.008 & 0.052:\tablenotemark{4}  \\
      & $-$0.003 & 0.009 & 1.018 & 0.021 & 0.035 & $-$0.023; & 0.017 & 1.064: & 0.040 & 0.075:  \\
\hline
$v-b$ & $-$0.002 & 0.001 & 0.987 & 0.003 & 0.020 & 0.019 & 0.001 & 1.001 & 0.003 & 0.011 \\
       & 0.024 & 0.009 & 0.961 & 0.008 & 0.019 & 0.027 & 0.006 & 0.982 & 0.005 & 0.022 \\
\hline
$u-v$ & $-$0.022 & 0.003 & 0.995 & 0.003 & 0.035 & $-$0.001: & 0.008 & 1.018: & 0.008 & 0.045: \\
      & $-$0.008 & 0.017 & 0.987 & 0.012 & 0.035 & 0.020: & 0.023 & 0.991: & 0.015 & 0.058: \\
\hline
$V-y$ & 0.000 & 0.000 & 0.011 & 0.001 & 0.002 & $-$0.004 & 0.000 & 0.011 & 0.001 & 0.004 \\
       & 0.009 & 0.004 & $-$0.009 & 0.007 & 0.003 & 0.012 & 0.004 & -0.160 & 0.006 & 0.011 \\
\enddata
\tablenotetext{1} {The synthetic photometry used zeropoint values determined from the CALSPEC stis005 Vega spectrum. The AB mag ZPs were $-$0.308, $-$0.327, $-$0.187 and $-$0.027 for $u$, $v$, $b$ and $y$, respectively.}
\tablenotetext{2} {It is recommended that the transformation equations derived from the NGSL stars be used to convert synthetic $uvby$ photometry onto the same system as the \citet{Hauc98} catalog. }
\tablenotetext{3} {The first line entry for each index is for $b-y< 0.5$. The second line is for $b-y>0.5$.}
\tablenotetext{4} {The $u$ and $c_{1}$ data for the MILES data is more uncertain because the spectra have been extrapolated 400\AA\ to the UV.}
\end{deluxetable}

\begin{deluxetable}{cccccccc}
\tabletypesize{\scriptsize}
\tablenum{3}
\tablecaption{Normalized \citep{Helt87} $uvby$ photonic passbands }
\tablewidth{0pt}
\tablehead{\colhead{Wave} & \colhead {$u$}&\colhead{Wave} & \colhead {$v$}& \colhead{Wave} & \colhead {$b$}& \colhead{Wave} & \colhead {$y$}}
\startdata
3320 & 0.000 & 4000 & 0.000 & 4560 & 0.000 & 5340 & 0.000 \\
3340 & 0.349 & 4020 & 0.427 & 4580 & 0.304 & 5360 & 0.389 \\
3360 & 0.492 & 4040 & 0.754 & 4600 & 0.576 & 5380 & 0.672 \\
3380 & 0.581 & 4060 & 0.884 & 4620 & 0.906 & 5400 & 0.912 \\
3400 & 0.687 & 4080 & 0.979 & 4640 & 0.994 & 5420 & 1.000 \\
3420 & 0.789 & 4100 & 0.995 & 4660 & 1.000 & 5440 & 0.982 \\
3440 & 0.835 & 4120 & 1.000 & 4680 & 0.969 & 5460 & 0.953 \\
3460 & 0.879 & 4140 & 0.996 & 4700 & 0.921 & 5480 & 0.913 \\
3480 & 0.926 & 4160 & 0.942 & 4720 & 0.871 & 5500 & 0.850 \\
3500 & 0.960 & 4180 & 0.742 & 4740 & 0.864 & 5520 & 0.798 \\
3520 & 0.982 & 4200 & 0.440 & 4760 & 0.804 & 5540 & 0.759 \\
3540 & 0.992 & 4220 & 0.219 & 4780 & 0.522 & 5560 & 0.725 \\
3560 & 1.000 & 4240 & 0.000 & 4800 & 0.186 & 5580 & 0.631 \\
3580 & 0.996 &           &  & 4820 & 0.000 & 5600 & 0.424 \\
3600 & 0.991 &  &  &                    &  & 5620 & 0.213 \\
3620 & 0.973 &  &  &  &                    & 5640 & 0.000 \\
3640 & 0.934 &  &  &  &  &  & \\
3660 & 0.816 &  &  &  &  &  & \\
3680 & 0.356 &  &  &  &  &  & \\
3700 & 0.000 &  &  &  &  &  & \\
\enddata
\end{deluxetable}

\begin{figure}
\figurenum{1}
\epsscale{.80}
\plotone{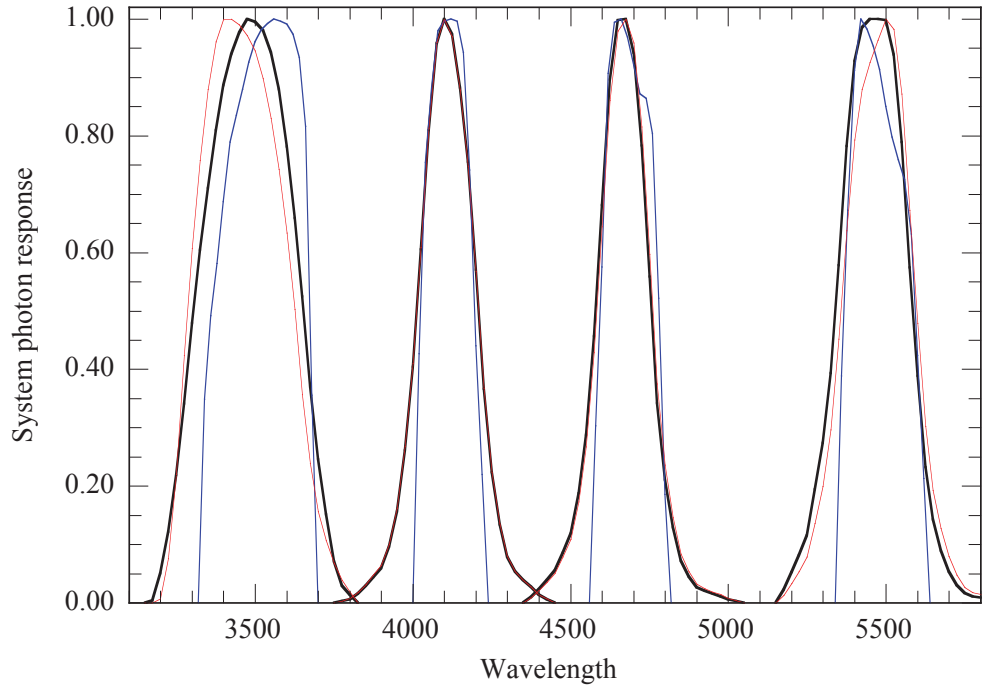}
\caption{The photonic passbands in this paper (black) compared with those of \citet{Maiz06}(red) and \citet{Helt87}(blue).}
\end{figure}

\begin{figure}
\figurenum{2}
\epsscale{.80}
\plotone{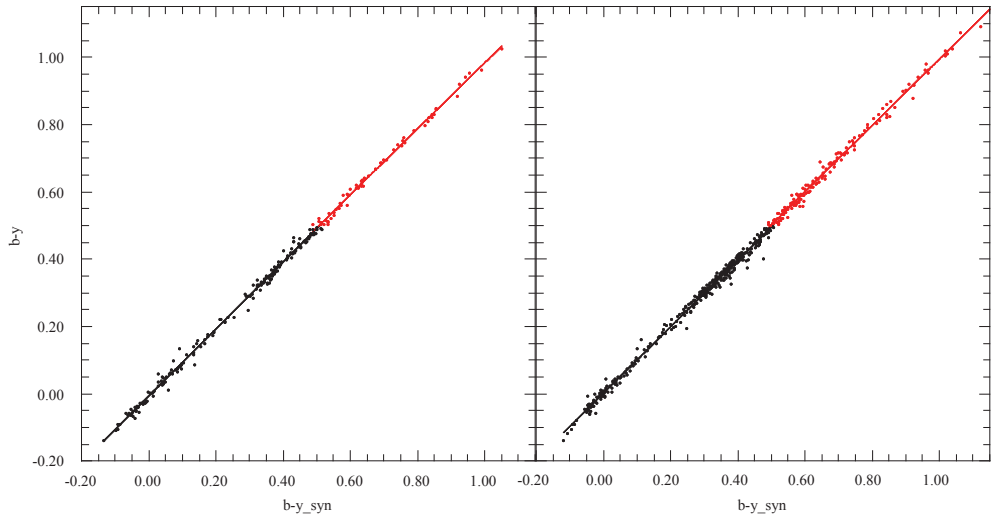}
\caption{The $b-y$ regressions. Stars with $b-y <$ 0.5 (black), $b-y >$ 0.5 (red). NGSL stars left; MILES stars right. The coefficients of the linear fits are given in Table 2.} 
\end{figure}

\begin{figure}
\figurenum{3}
\epsscale{.80}
\plotone{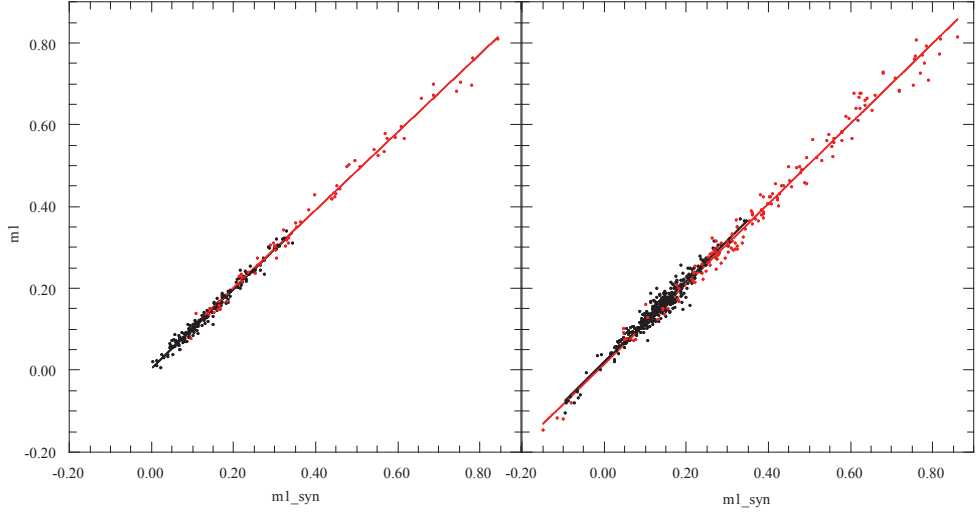}
\caption{The $m_{1}$ regressions. } 
\end{figure}

\begin{figure}
\figurenum{4}
\epsscale{.80}
\plotone{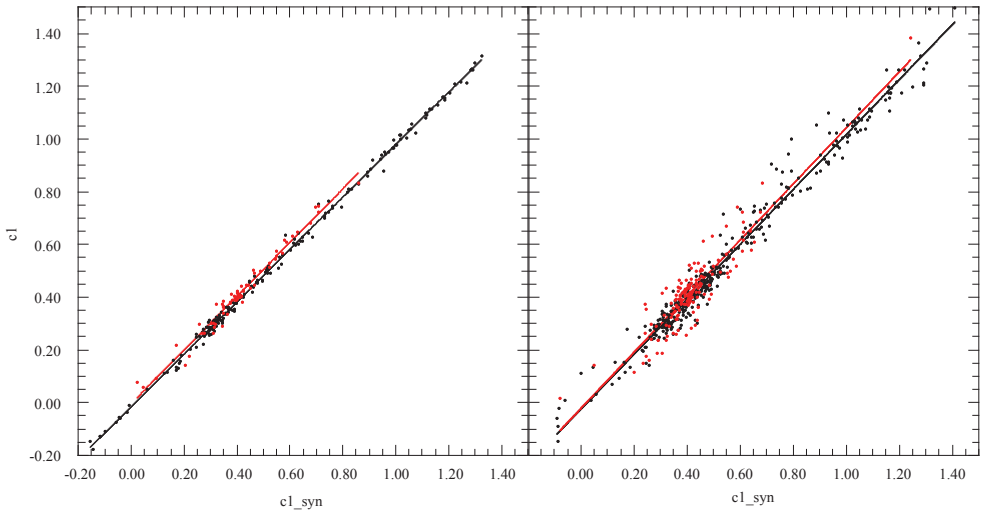}
\caption{The $c_{1}$ regressions.} 
\end{figure}

\begin{figure}
\figurenum{5}
\epsscale{.80}
\plotone{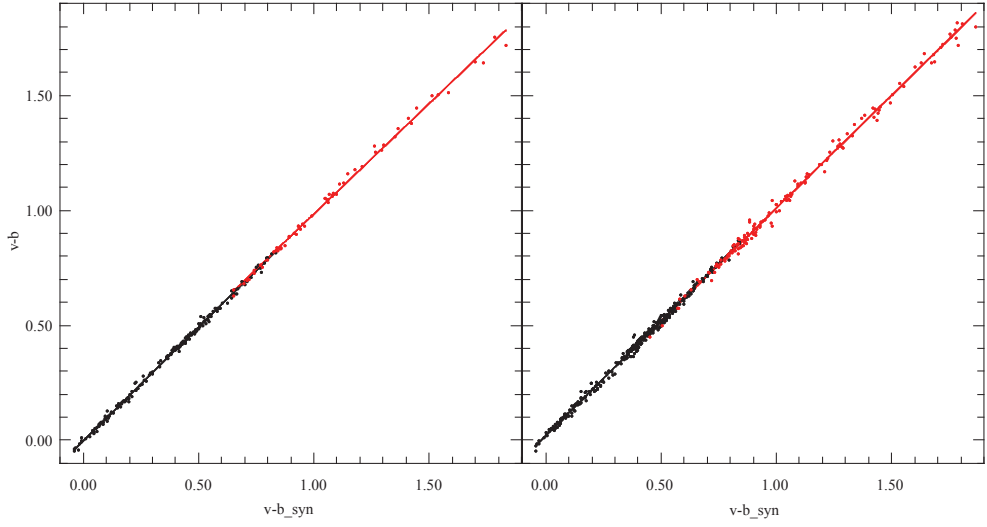}
\caption{The $v-b$ regressions.} 
\end{figure}

\begin{figure}
\figurenum{6}
\epsscale{.80}
\plotone{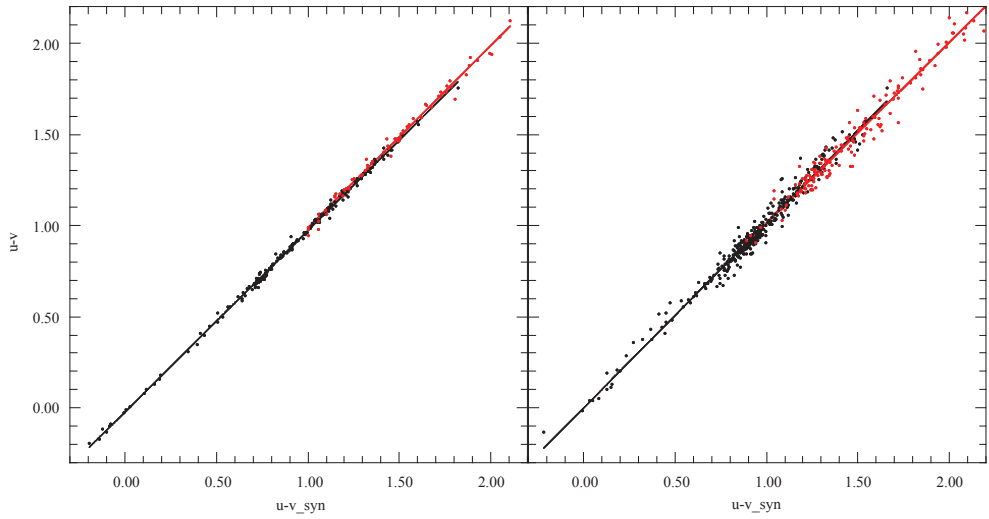}
\caption{The $u-v$ regressions.} 
\end{figure}

\begin{figure}
\figurenum{7}
\epsscale{.80}
\plotone{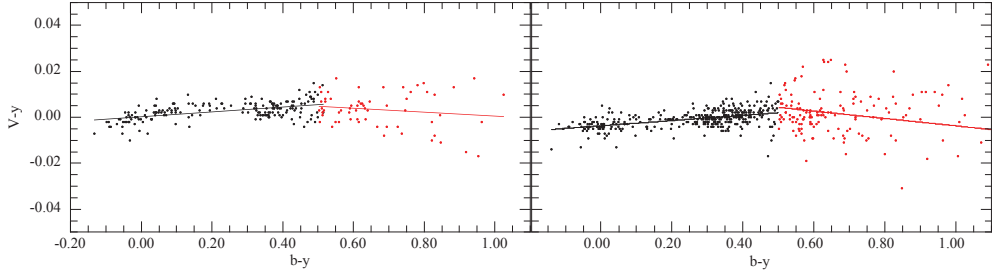}
\caption{The synthetic $V-y$ regressions.} 
\end{figure}

\begin{figure}
\figurenum{8}
\epsscale{.80}
\plotone{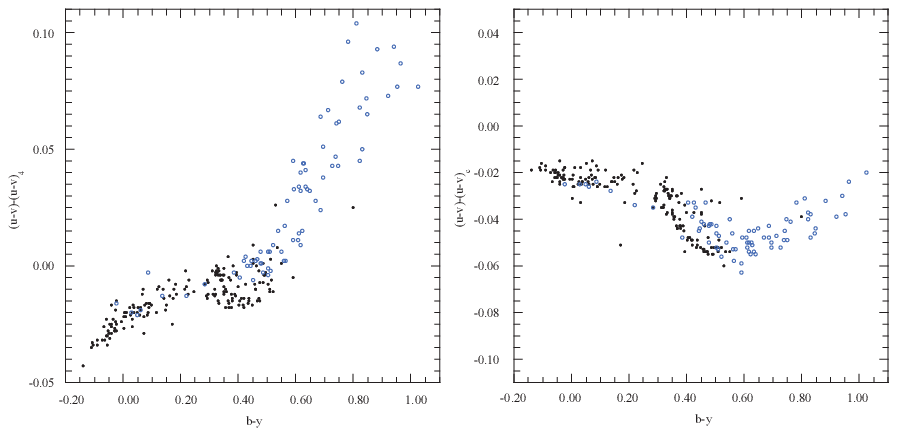}
\caption{The synthetic $\Delta$ ($u-v$) regressions for the natural 4-channel system \citep{Helt87} (left) and \citet{Cous87} (right). Dwarfs (black dots), giants (blue open circles)} 
\end{figure}

\begin{figure}
\figurenum{9}
\epsscale{.80}
\plotone{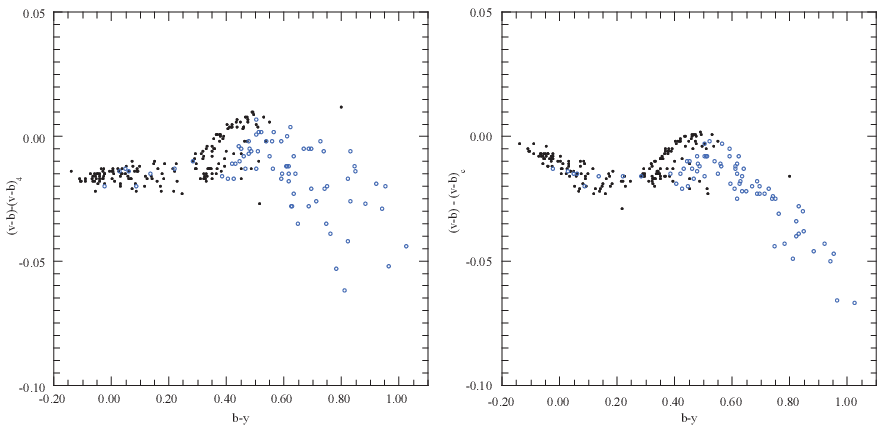}
\caption{The synthetic $\Delta$ ($v-b$) regressions for the natural 4-channel system \citep{Helt87} (left) and \citet{Cous87} (right). Dwarfs (black dots), giants (blue open circles)} 
\end{figure}

\begin{figure}
\figurenum{10}
\epsscale{.80}
\plotone{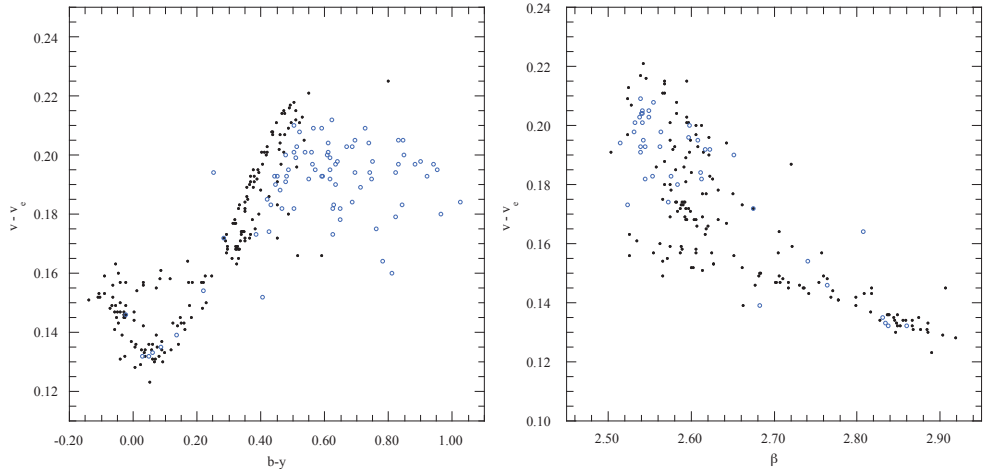}
\caption{Synthetic $\Delta$ $v$ versus (b-y) and $\beta$ for the narrow $v$ filter. Dwarfs (black dots), giants (blue open circles) } 
\end{figure}

\end{document}